\documentclass[preprint]{aastex}

\usepackage{amsmath}
\usepackage{amssymb}
\usepackage{bm}

\slugcomment{accepted for publication in ApJ}

\shorttitle{}
\shortauthors{Odaka et al.}

\begin{document}


\title{X-ray Diagnostics of Giant Molecular Clouds in the Galactic Center Region and Past Activity of Sgr A*}


\author{Hirokazu Odaka\altaffilmark{1,2}, Felix Aharonian\altaffilmark{3,4}, Shin Watanabe\altaffilmark{1}, Yasuyuki Tanaka\altaffilmark{1},\\ Dmitry Khangulyan\altaffilmark{1,4}, and Tadayuki Takahashi\altaffilmark{1,2}}


\altaffiltext{1}{ Institute of Space and Astronautical Science (ISAS), Japan Aerospace Exploration Agency (JAXA), 3-1-1 Yoshinodai, Sagamihara, Kanagawa, 252-5210, Japan}
\altaffiltext{2}{Department of Physics, University of Tokyo, 7-3-1 Hongo, Bunkyo, Tokyo, 113-0033, Japan}
\altaffiltext{3}{Dublin Institute for Advanced Studies (DIAS), 31 Fitzwilliam Place, Dublin 2, Ireland}
\altaffiltext{4}{Max-Planck-Institut f\"ur Kernphysik, Saupfercheckweg 1, Heidelberg 69117, Germany}


\begin{abstract}
Strong iron fluorescence at 6.4 keV and hard-X-ray emissions from giant molecular clouds in the Galactic center region have been interpreted as reflections of a past outburst of the Sgr A* supermassive black hole.
Careful treatment of multiple interactions of photons in a complicated geometry is essential to modeling the reprocessed emissions from the dense clouds.
We develop a new calculation framework of X-ray reflection from molecular clouds based on Monte Carlo simulations for accurate interpretation of high-quality observational data.
By utilizing this simulation framework, we present the first calculations of morphologies and spectra of the reflected X-ray emission for several realistic models of Sgr B2, which is the most massive molecular cloud in our Galaxy.
The morphology of scattered hard X-rays above 20 keV is significantly different from that of iron fluorescence due to their large penetrating power into dense regions of the cloud, probing the structure of the cloud.
High-resolution spectra provide quantitative evaluation of the iron line including its Compton shoulder to constrain the mass and the chemical composition of the cloud as well as the luminosity of the illuminating source.
These predictions can be checked in the near future with future X-ray missions such as {\it NuStar} (hard X-rays) and {\it ASTRO-H} (both iron lines and hard X-rays).
\end{abstract}




\bibliographystyle{apj}


\section{Introduction}

Sgr A*, the supermassive black hole with a mass of $4\times 10^6M_\odot$ at the center of our Galaxy \citep{Schodel:2002, Ghez:2003, Gillessen:2009}, is currently very faint with a luminosity of $10^{33}$-$10^{34}\ \mathrm{erg\ s^{-1}}$ \citep{Baganoff:2001}, which is many orders of magnitude lower than that of active galactic nuclei (AGNs).
Such quiescent activity of the nucleus seems inconsistent with the dense environment in the Galactic center  (GC) from which the black hole could accrete sufficient matter \citep{Melia:2001}.
Although Sgr A* displays flares whose luminosity increases by a factor of $\sim$100 \citep{Baganoff:2001}, the peak luminosity is still $10^8$ times dimmer than the Eddington luminosity of the supermassive black hole.
Whether Sgr A* has always been so quiescent or previously experienced high activity such as low-luminosity AGNs is an important question in addressing the evolution and structure of the Galaxy or the accretion mechanism taking place at the center of the Galaxy.

Indications of past activities of Sgr A* have come from X-ray observations of giant molecular clouds in the GC region.
\citet{Sunyaev:1993} interpreted X-ray emissions associated with molecular clouds as reflections of a past outburst of Sgr A* and predicted a correlation of 6.4 keV iron K$\alpha$ fluorescent line emission with molecular clouds.
\citet{Koyama:1996} discovered with {\it ASCA} strong 6.4 keV emissions associated with the Sgr B2 giant molecular cloud, which is located at a projected distance of 100 pc from Sgr A*, and indicated that the cloud is possibly irradiated by a flare of Sgr A* with a luminosity of $2\times 10^{39}\ \mathrm{erg\ s^{-1}}$ which occurred 300 years ago.
Against this X-ray reflection nebulae (XRNe) interpretation, the iron fluorescence could be generated by collisions with cosmic-ray electrons \citep{YusefZadeh:2007} or protons \citep{Dogiel:2009}.
In order to understand the origin of the fluorescence, time variability of the emission is essential since rapid variability over a few years in the region of a parsec scale is hard to explain by cosmic rays.
In actual fact, detections of time variability of the iron fluorescent line from Sgr B2 \citep{Koyama:2008, Inui:2009} and molecular clouds around Sgr A \citep{Muno:2007, Ponti:2010} in a timescale of a few years have provided strong support for the XRN model.

In the picture of the XRN model, since X-ray photons from the Sgr A* flare are reprocessed via the photoelectric effect or scattered in the dense clouds, the observed X-ray emissions from the clouds contain important information about the flare and the molecular clouds.
From this viewpoint, \citet{Sunyaev:1998} provided theoretical calculations about iron fluorescence emissions from an illuminated cloud.
Based on a simplified model, they have realized several general effects on time evolution of the spectral and morphological characteristics of the reflected X-ray emission.
They also pointed out that a low-energy tail structure, or a Compton shoulder, of the iron line, which will be investigable by future high-resolution imaging spectroscopy, contains information on the time elapsed since the outburst.
The strong iron fluorescence is the most prominent feature of XRNe and has great potential for detailed measurements of giant molecular clouds in the GC region.

In addition to the iron 6.4 keV lines, hard X-rays scattered via Compton scattering above 12 keV from the clouds also have good diagnostic power.
Above 12 keV, the cross section of photoelectric absorption is smaller than the Thomson cross section for the chemical composition identical with the solar abundances.
Since the hard X-rays are less affected by absorption than soft X-rays, they reflect the structure of dense regions in the clouds directly.
\citet{Revnivtsev:2004} reported hard X-rays up to $\sim$200 keV associated with the Sgr B2 molecular cloud observed with {\it International Gamma-Ray Astrophysics Laboratory} ({\it INTEGRAL}).
Moreover, continuous monitoring of Sgr B2 with {\it INTEGRAL} for seven years revealed fading of the hard X-ray emission whose decay rate is consistent with the iron line variability \citep{Terrier:2010}.
However, its morphology has not yet been obtained due to insufficient angular resolutions and high background levels of current hard-X-ray observations.

Future X-ray missions in the next decade will provide new observational information on XRNe in the GC region.
{\it ASTRO-H} \citep{Takahashi2010}, which is the next international X-ray satellite scheduled for launch in 2014, will simultaneously realize both high-resolution imaging spectroscopy with an excellent energy resolution of 7 eV (full width at half maximum or FWHM) at 6.4 keV and hard-X-ray focused imaging up to 80 keV.
The unprecedented spectral resolution of the soft X-ray spectrometer system (SXS) onboard {\it ASTRO-H} will enable us to resolve the fine structure of the iron line complex including the Compton shoulder.
The {\it NuStar} mission \citep{Harrison2010} will also open a window of hard-X-ray imaging, scheduled for launch in 2012.
In 2020's {\it International X-ray Observatory (IXO)} \citep{Bookbinder2010} will be able to observe the molecular clouds with both high spectral and high spatial resolution.

In order to interpret high-quality observations with such future missions, a high-quality model of the X-ray emission from the illuminated clouds is also required.
Since giant molecular clouds in the GC region are very massive ($\sim 10^{5}-10^{6}~M_\odot$) and the Thomson optical depth reaches $\sim$1, absorption by heavy elements and scattering by hydrogen molecules and helium atoms in a complex geometry should be treated with care.
For detailed treatment of a complicated geometry of the clouds, \citet{Murakami:2000} developed a numerical model of XRNe to constrain physical properties of the clouds by comparing the numerical calculations to observations.
This approach, however, would be insufficient for high-resolution data due to difficulties with multiple interactions.
Multiple scattering is essential for modeling the Compton shoulder and scattered hard X-rays in such dense clouds.
Another approach is adoption of Monte Carlo simulations to solve the problem of radiative transfer \citep{Leahy:1993, Sunyaev:1998, Fromerth:2001}.
In general, Monte Carlo simulations treat particle tracking by calculating the propagation and interactions of a photon in matter.
Although the Monte Carlo simulation is suitable for multiple interactions, calculations of illuminated clouds have been limited to a problem of a simple geometry with homogeneous matter distribution.
Therefore, a Monte Carlo simulation that treats both multiple interactions and a complicated geometry is necessary for accurate calculation of the XRN emissions.

In this work, we have constructed a new model describing the X-ray emission from the clouds based on the Monte Carlo simulation.
Our approach appropriately treats multiple interactions of photons in a realistic cloud geometry after a technique of X-ray spectral simulation of high-mass X-ray binaries developed by \citet{Watanabe:2003, Watanabe:2006}.
Based on this model, we have investigated prospective diagnostics of giant molecular clouds in the GC region with high-resolution spectroscopy of an iron fluorescent line at 6.4 keV and hard-X-ray imaging.
The simulation we developed provides the first calculation of X-ray images and spectra from an XRN which has a realistic configuration, aimed at practical observations.
The future X-ray missions will be able to perform the diagnostic observations proposed in this paper with the help of our calculation framework.
The present paper describes design of the simulation (\S\ref{subsec:monte_carlo}), simulated geometry models of molecular clouds (\S\ref{subsec:geometry}), results of the simulation (\S\ref{sec:results}), and discussion on the results and observational prospects for the future missions (\S\ref{sec:discussion}).

\section{Methodology}\label{sec:methodology}

\subsection{Monte Carlo Simulation}\label{subsec:monte_carlo}

In order to calculate X-ray emissions from illuminated clouds, we constructed a new framework based on Monte Carlo simulation.
This framework, which is called MONACO (standing for MONte Carlo simulation for Astrophysics and COsmology), is a general-purpose code applicable to photon-tracking simulations of many astrophysical objects.
We utilize the Geant4 simulation toolkit library \citep{Agostinelli:2003, Allison:2006} for tracking particles and building geometry, while physical processes are our original implementations.
Geant4, which is widely used for experimental particle physics and nuclear science, provides sophisticated handling of complicated geometries such as radiation detectors.

\begin{figure}[htbp]
\begin{center}
\includegraphics[width=8cm]{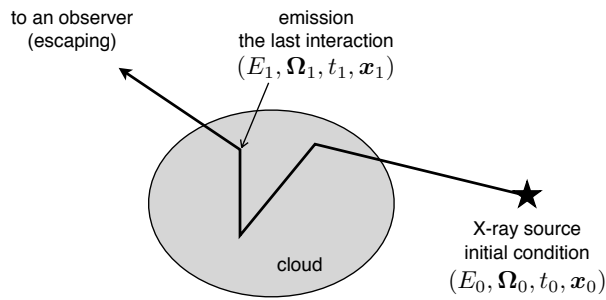}
\caption{Schematic concept of the Monte Carlo simulation.}
\label{fig:concept}
\end{center}
\end{figure}

Figure~\ref{fig:concept} shows a schematic concept of the Monte Carlo simulation.
In the simulation, we build a cloud geometry system containing physical properties such as chemical composition and number density of the cloud; then in each Monte Carlo trial, a photon with energy $E_0$ and direction $\bm{\Omega}_0$ is generated at time $t_0$ and position $\bm{x}_0$.
The simulation tracks the particle until it escapes from the system, recording information about the last interaction of the photon $(E_1, \bm{\Omega_1}, t_1, \bm{x}_1)$, which an observer regards as emission.
A detailed description of the Monte Carlo simulation and its framework is given in Appendix \ref{app:framework}.

% \subsection{Physical processes}\label{subsec:physical_process}

We have developed our original implementation of physical processes adaptable to Monte Carlo simulation frameworks based on Geant4. 
In neutral matter such as molecular clouds, relevant interactions of X-rays are photoelectric absorption and scattering by bound electrons in hydrogen and helium.
Scattering by electrons in heavy elements is negligible because of their small abundances.
While the simulation tracks a photon by Monte Carlo methods, an interaction occurs according to its mean free path; then a status of the photon is changed according to differential cross section of the interaction.

As cross section data of photoelectric absorption, we adopt data based on the Evaluated Photon Data Library 97 (EPDL97)\footnote{http://www-nds.iaea.org/epdl97/}, which is provided along with the Geant4 toolkit as the data of Low-Energy Electromagnetic processes version 6.9.
A K-shell fluorescent photon is generated with a probability of a fluorescence yield, following a photoelectric effect.
As relevant atomic properties, line energies of K-shell fluorescence, K-shell fluorescence yields, and K$\beta$-to-K$\alpha$ ratios are taken from \citet{Thompson2001}, \citet{Krause1979}, and \citet{Ertugral2007}, respectively.
In addition, K$\alpha_2$-to-K$\alpha_1$ ratios are fixed to 0.5.

Scattering by bound electrons can be divided into three channels according to the final state of the target electron: Rayleigh scattering (elastic), Raman scattering (excitation), and Compton scattering (ionization).
The difference in the target between a free electron and a bound electron is significant for the fine structure of the iron line complex including the Compton shoulder.
The shape of the Compton shoulder is smeared by an electron binding effect which is known as the Doppler broadening effect due to the finite momentum of the electron bound to an atom or molecule.
For atomic hydrogen, cross sections and differential cross sections of the three channels are provided by analytical formulae \citep{Sunyaev:1996}.
For molecular hydrogen, we adopt an approximation that scattering by the molecule is identical to that by atomic hydrogen except that the cross section of Rayleigh scattering is enhanced by a factor of two per electron \citep{Sunyaev:1999}.
Since the second most abundant element, helium, does not have analytic expressions of atomic properties, we use numerical calculations of differential cross sections \citep{Vainshtein:1998}.
Appendix \ref{app:process_scattering} provides details of the implementation of the scattering processes.

\begin{figure}[htbp]
\begin{center}
\includegraphics[width=7cm]{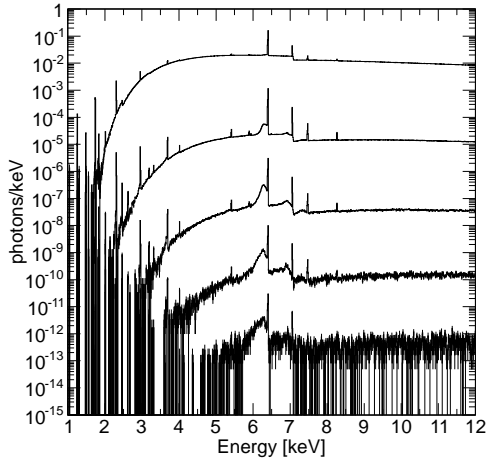}
\caption{Spectra emerging from a spherical uniform cloud with a radial Thomson depth of 0.2 in which a short flare occurred at $t=0$. The spectra are extracted in different time intervals of $0<t<R/c$, $R/c<t<2R/c$, $2R/c<t<3R/c$, $3R/c<t<4R/c$, and $4R/c<t<5R/c$ (from top to bottom), where $R$ is the cloud radius. Each subsequent spectrum is multiplied by 0.05 for clarity. This simulation setup is identical to the conditions of Fig.~9 of Sunyaev \& Churazov (1998).}
\label{fig:verification}
\end{center}
\end{figure}

For verification of our simulation, we performed a benchmark calculation of an illuminated cloud.
Results of this simulation can be directly compared with Fig.~9 of Sunyaev \& Churazov (1998) under identical simulation conditions.
In the simulation setup, a point X-ray source was located at the center of a spherical uniform cloud that has a radial Thomson depth of 0.2 from the center to the surface.
Figure~\ref{fig:verification} shows emerging spectra from the cloud in different time intervals when a short flare occurred at $t=0$.
The initial spectrum was a power law with a photon index of 1.8 in an energy range of 1--18 keV.
Our results nicely agree with the former calculation on the wide-band continuum and the iron K$\alpha$ line shape including the Compton shoulder for each time interval.
In addition, the spectra from our calculation present iron K$\beta$ at 7.1 keV and K-shell fluorescent lines of other metals, which were neglected in Sunyaev \& Churazov (1998).

\subsection{Model of Sgr B2}\label{subsec:geometry}

We used the simulation framework described in \S\ref{subsec:monte_carlo} to model X-ray reflection from the Sgr B2 giant molecular cloud in the GC region.
Sgr B2 is one of the brightest XRNe emitting strong 6.4 keV iron fluorescence, which also shows variability over a timescale of a few years.
Moreover, since this cloud has been intensively studied in other wavelengths such as radio and infrared bands, its structure and physical properties such as velocity fields and density distribution of molecular gas are relatively well measured compared to other molecular clouds.
Sgr B2 is, therefore, the best target for investigating X-ray reflection from molecular clouds.

\begin{figure}[htbp]
\begin{center}
\includegraphics[width=7cm]{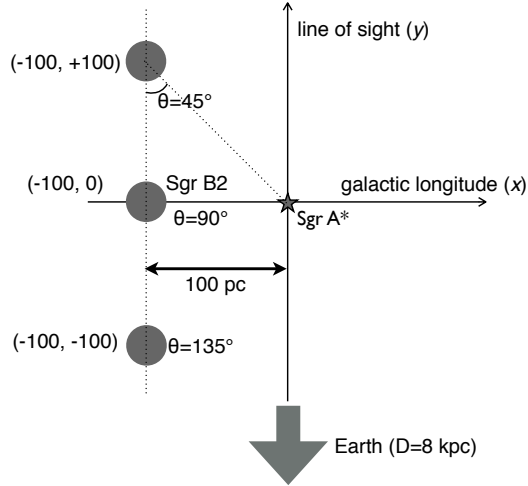}
\caption{Geometrical setup of the simulation. We assumed three different positions of the Sgr B2 cloud, fixing the projected distance from Sgr A* to 100 pc.}
\label{fig:gc_map}
\end{center}
\end{figure}

Figure~\ref{fig:gc_map} shows the geometrical setup of the simulation.
A photon generator or X-ray source simulating the Sgr A* supermassive black hole is located at the center.
The projected distance between Sgr B2 and Sgr A* is fixed to 100 pc.
Although a trigonometric parallax of Sgr B2 with the Very Long Base Array \citep{Reid:2009} and X-ray absorption measurements toward the GC region with the {\it Suzaku} satellite \citep{Ryu:2009} suggested that Sgr B2 is positioned in front of the GC, there still remains uncertainty of the position of the cloud along the line of sight.
The line-of-sight position is important in studying the structure of the Galaxy and also makes a difference in the time of the Sgr A* outburst.
We assumed three different line-of-sight positions of the cloud; $y=+100\ \mathrm{pc}$, $0\ \mathrm{pc}$, and $-100\ \mathrm{pc}$, to investigate effects of the relative position of the cloud to the center.
In the present paper, we assume that the distance to the GC from the Earth is 8 kpc.

The Sgr B2 complex is the most massive giant molecular cloud in our Galaxy and has a complicated structure containing several active star forming HII regions.
As intensive observations in radio bands have provided models of the cloud's structure \citep{Lis:1990, Hasegawa:1994, deVicente:1997}, it can be presumed that the cloud consists of three components: (1) very dense cores with diameters of $\sim 0.5\ \mathrm{pc}$ containing star-forming regions, (2) a dense envelope which seems to correspond with an XRN with a number density of hydrogen molecule of $n_\mathrm{H_2}=10^{4}$-$10^{5} \ \mathrm{cm^{-3}}$, and (3) a large, diffuse component with an almost constant density of $n_\mathrm{H_2}\sim 10^{3}\ \mathrm{cm^{-3}}$ extending to a diameter of 45 pc.

For simplicity we constructed a spherically symmetric geometry model composed of a single dense core and an envelope that has a diameter of 10 pc and a realistic density distribution.
We determined the diameter of the envelope according to Hasegawa et al.\ (1994) and the value well agrees with X-ray morphology obtained by {\it Chandra} observations \citep{Murakami:2001}.
In the model the radial profile of the number density in the cloud is represented as
\begin{equation}
n(r)=\left\{ \begin{array}{lll}
5\times 10^6\ \mathrm{cm}^{-3} & (\text{core: }r<0.25\ \mathrm{pc}) \\
n_0 \left(\dfrac{r}{r_0}\right)^{-\alpha} \mathrm{cm}^{-3} & (\text{envelope: }0.25\ \mathrm{pc}<r<5.0\ \mathrm{pc}) 
\end{array}
\right.
\end{equation}
The mass of the core is $2\times 10^4\ M_\odot$, which is a very small fraction of the total mass of the cloud.
Table~\ref{table:model_param} shows parameters of our cloud models in the simulation.
We prepared four models with different $n_0$ values corresponding to total cloud masses from $2.5\times 10^5M_\odot$ to $2\times 10^6M_\odot$ (Models 1--4).
In Models 1--4, the power-law index $\alpha=1$ was assumed, as radio observations have reported values of 2 \citep{Lis:1990} or 0.87 \citep{deVicente:1997}.
To investigate effects of different density profiles, we built a model with $\alpha=0$ (Model 5) and a model with $\alpha=2$ (Model 6), fixing the total mass to $5\times 10^5 M_\odot$.
While we assumed a metal abundance of 1.5 protosolar value in Models 1--6 as a standard value in the GC region \citep{Nobukawa:2010}, we additionally checked two models of different values of 1.0 protosolar (Model 7) and 2.0 protosolar (Model 8).
The values of the protosolar (solar system) abundances are taken from \citet{Lodders:2003}.
We ignored the third component surrounding the dense envelope to reduce computation costs; instead we introduced absorption of the initial spectrum by the third component as
\begin{equation}
F(E)\propto \exp(-N_\mathrm{H}\sigma_\text{abs}(E))E^{-\gamma}\quad (1\ \mathrm{keV}<E< 400\ \mathrm{keV}),
\end{equation}
where $N_\mathrm{H}=6\times 10^{22}\ \mathrm{cm}^{-2}$ is an equivalent hydrogen column density of the surrounding diffuse component, and $\sigma_\text{abs}(E)$ is the photoelectric absorption cross section at energy $E$ per hydrogen.
The photon index $\gamma$ of the initial spectrum is 1.8, which is consistent with flares of Sgr A* or X-ray emission from Seyfert galaxies.

\begin{table}[htdp]
\caption{Parameters of the cloud models: power-law index $\alpha$ of the radial density profile of the cloud envelope, the normalization of the power law, the metal abundance of the cloud, and the total mass of the cloud calculated by these parameters.}
\begin{center}
\begin{tabular}{ccccc}
\hline\hline 
\# & $\alpha$ & $n_0$ & metal abundance & $M_\text{cloud}$ \\
 &  & [$10^4\ \mathrm{cm^{-3}}$] & [protosolar] & [$10^5M_\odot$] \\
\hline
1 & 1.0 & 2.5 & 1.5 & 2.5 \\
2 & 1.0 & 5.0 & 1.5 & 5.0 \\
3 & 1.0 & 10.0 & 1.5 & 10.0 \\
4 & 1.0 & 20.0 & 1.5 & 20.0\\
\hline
5 & 0.0 & 1.5 & 1.5 & 5.0 \\
6 & 2.0 & 13.1 & 1.5 & 5.0 \\
\hline
7 & 1.0 & 5.0 & 1.0 & 5.0 \\
8 & 1.0 & 5.0 & 2.0 & 5.0 \\
\hline
\end{tabular}
\end{center}
\label{table:model_param}
\end{table}%

\section{Results}\label{sec:results}

\subsection{Morphology}\label{subsec:morphology}

We investigated morphologies of iron 6.4 keV line emissions and hard X-ray emissions by extracting images from the simulation results.
Since X-ray emissions extend over the whole cloud, a duration of the illuminating flare of Sgr A* must be longer than the light crossing time of the cloud.
As the exact value of the duration does not significantly affect the emissions unless it is shorter than the light crossing time, we assumed that the flare began at $t=-65$ yr, kept a constant luminosity, and then ended at $t=0$.
The duration of 65 years is twice as long as the light crossing time of the 10 pc large cloud.
The X-ray reflection from the cloud becomes brightest just before the rear end of the direct light front reaches the cloud edge confronting the illuminating source.
Time of the brightest moment depends on the relative position of the cloud to the flare source and the observer, as summarized in Table~\ref{table:brightest}.
The assumed luminosity of the flare is $L_X=1.3\times 10^{39}\ (d/100\ \mathrm{pc})^2 \ \mathrm{erg\ s^{-1}}$ in an energy range of 1--10 keV, where $d=\sqrt{x^2+y^2}$ is the distance between the cloud and the illuminating source.
This is normalized so that the iron line flux of Model 2 ($y=0$ pc) at the brightest moment agrees with an observational value of $1.0\times 10^{-4}$ photons $\mathrm{s}^{-1}\mathrm{cm}^{-2}$ obtained with {\it Chandra} in 2000 \citep{Inui:2009}.

\begin{table}[htdp]
\caption{Time of the brightest moment elapsed since the end of the flare ($t=0$) for different cloud positions.}
\begin{center}
\begin{tabular}{cc}
\hline\hline 
position & time of the brightest moment \\
$[\mathrm{pc}]$ & $[\mathrm{yr}]$ \\
\hline
$(-100,\ -100)$ & 122 \\
$(-100,\ 0)$ & 303 \\
$(-100,\ +100)$ & 756 \\
\hline
\end{tabular}
\end{center}
\label{table:brightest}
\end{table}%

\begin{figure}[htbp]
\begin{center}
\includegraphics[width=16.5cm]{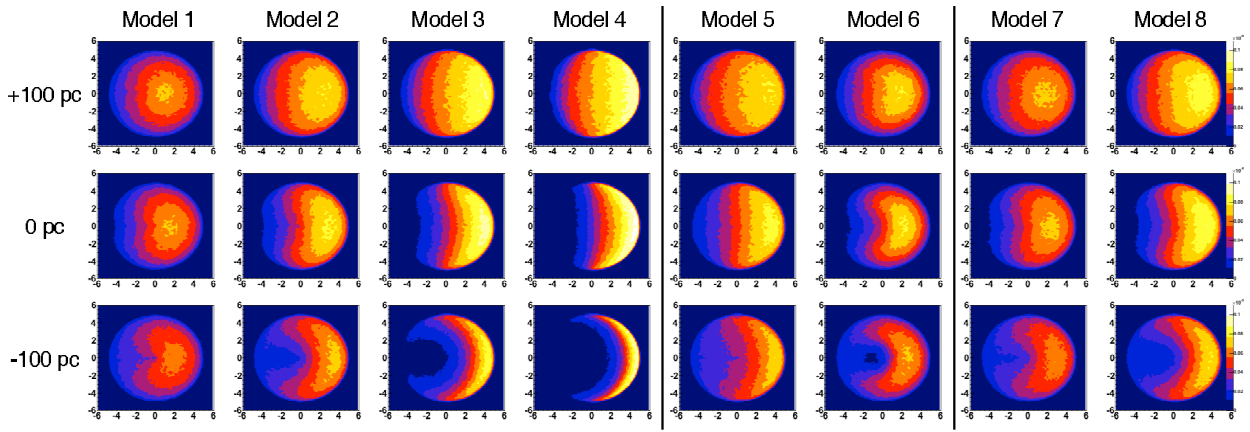}
\caption{Morphologies of iron K$\alpha$ 6.4 keV line emissions for Models 1--8 at the brightest moment. Each row corresponds to a different line-of-sight position of the cloud marked at the left. The colors are mapped in linear scale and are common in all the images. The size of each square image is 12~pc$\times$12~pc, corresponding to an angular size of $5'.2 \times 5'.2$.}
\label{fig:image_iron_model}
\end{center}
\end{figure}

\begin{figure}[htbp]
\begin{center}
\includegraphics[width=16.5cm]{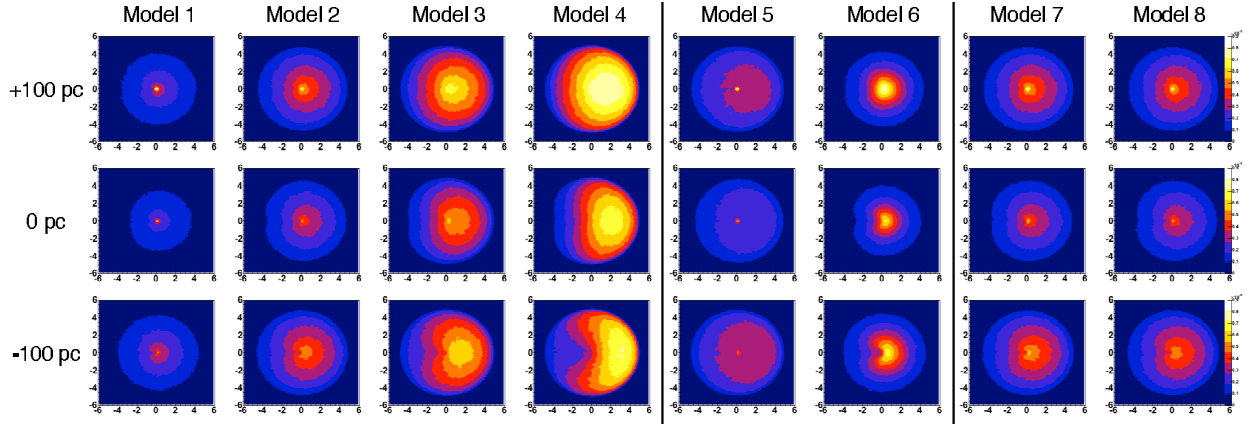}
\caption{Same as Fig.~\ref{fig:image_iron_model} but for hard X-ray emissions in an energy range of 20--60 keV.}
\label{fig:image_hard_model}
\end{center}
\end{figure}

Figure~\ref{fig:image_iron_model} depicts iron 6.4 keV line morphologies at the brightest moment for the eight different cloud models and three different line-of-sight positions.
The X-ray illuminates the cloud from the right side of each image.
The case of the middle position ($y=0$ pc) is shown in the middle row of Fig.~\ref{fig:image_iron_model}.
In this case, the cloud becomes brightest at $t=303$ yr.
As the cloud mass increases (from Model 1 to 4), the cloud becomes opaque, making the morphology semicircular.
Corresponding hard-X-ray images extracted in an energy range between 20 keV and 60 keV are shown in the middle row of Fig.~\ref{fig:image_hard_model}.
The hard-X-ray morphologies are obviously different from the iron fluorescence due to their large penetrating power.
Hard X-rays are therefore useful for probing the structure of the dense molecular clouds.
Images of Models 5, 2, and 6 show the difference due to the density profile of the cloud.
Since the cloud with a steeper radial profile (Model 6) has denser inner layers and thinner outer layers, its morphology becomes more concentrated and concave.
In Model 5, the small opacity of the uniform envelope exposes the central core in the hard X-ray band.
The metal abundance in the cloud also changes the iron line emission while the hard-X-ray images are similar.

If the line-of-sight position of the cloud is different, the incident direction of the illuminating photons changes and significantly different morphologies appear.
The top rows in Figures \ref{fig:image_iron_model} and \ref{fig:image_hard_model} show X-ray morphologies at the brightest moment, $t=756\ \mathrm{yr}$, in the  front-illuminated case or the position of $y=+100\ \mathrm{pc}$.
The delay of the brightest moment is due to light propagation via the cloud behind the GC.
In this case, the iron line emissions form into a round shape since a large proportion of the cloud front is illuminated by the flare.
On the other hand, the bottom row of Figures \ref{fig:image_iron_model} and \ref{fig:image_hard_model} show morphologies in the back-illuminated case or the position of $y=-100\ \mathrm{pc}$, and the reflection becomes the brightest at $t=122\ \mathrm{yr}$.
Morphologies of the iron line form into a concave shape; moreover, the shape becomes a crescent if the cloud is more massive than $\sim 1\times 10^6M_\odot$.
In contrast to the iron emission, morphologies of scattered hard X-rays do not largely depend on the relative position of the cloud or the direction of the illuminating source unless the cloud is Compton thick.

\begin{figure}[htbp]
\begin{center}
\includegraphics[width=16.5cm]{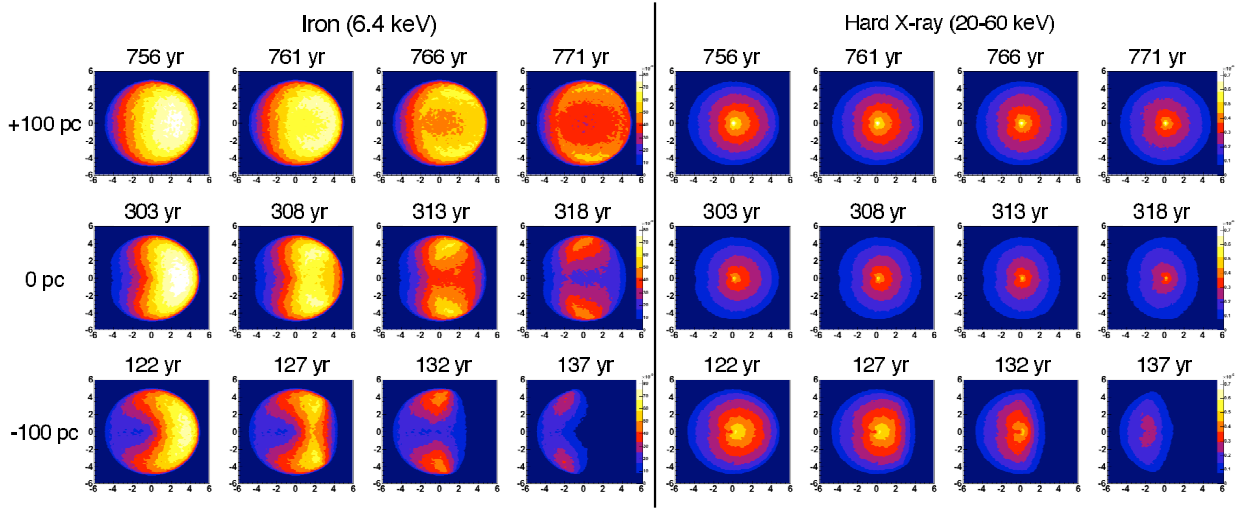}
\caption{Time evolution of the morphology of the iron line (left panels) and the hard X-ray (right panels) for Model 2 from the brightest moment at intervals of five years. The observation time is marked at the top of each image and the cloud position along the line of sight is marked at the left of each row. The colors are mapped in linear scale and are common in all the images of the same energy band. }
\label{fig:image_time}
\end{center}
\end{figure}

Since Sgr B2 displays fading of the X-ray emission \citep{Koyama:2008, Inui:2009, Terrier:2010}, we would observe evolution of the X-ray morphologies in the XRN model as shown in Fig.~\ref{fig:image_time}.
At 6.4 keV the central region of the cloud will become darker in the fading phase than outer regions if the cloud is massive enough to absorb most of the incident soft-X-ray photons in the central regions.
In contrast, hard X-rays do not suffer from such significant absorption; thus morphological disagreement between the iron line and the hard X-ray is notable in the decay phase.
An apparent light front intersecting the cloud from the observer moves fast when the cloud is positioned in front of the flare source \citep{Sunyaev:1998}.
For the cloud at $y=-100\ \mathrm{pc}$ morphological evolution is very fast even in the hard X-ray band (Fig.~\ref{fig:image_time} bottom).

\subsection{Spectral Properties}\label{subsec:spectra}

\begin{figure}[htbp]
\begin{center}
\includegraphics[width=8.5cm]{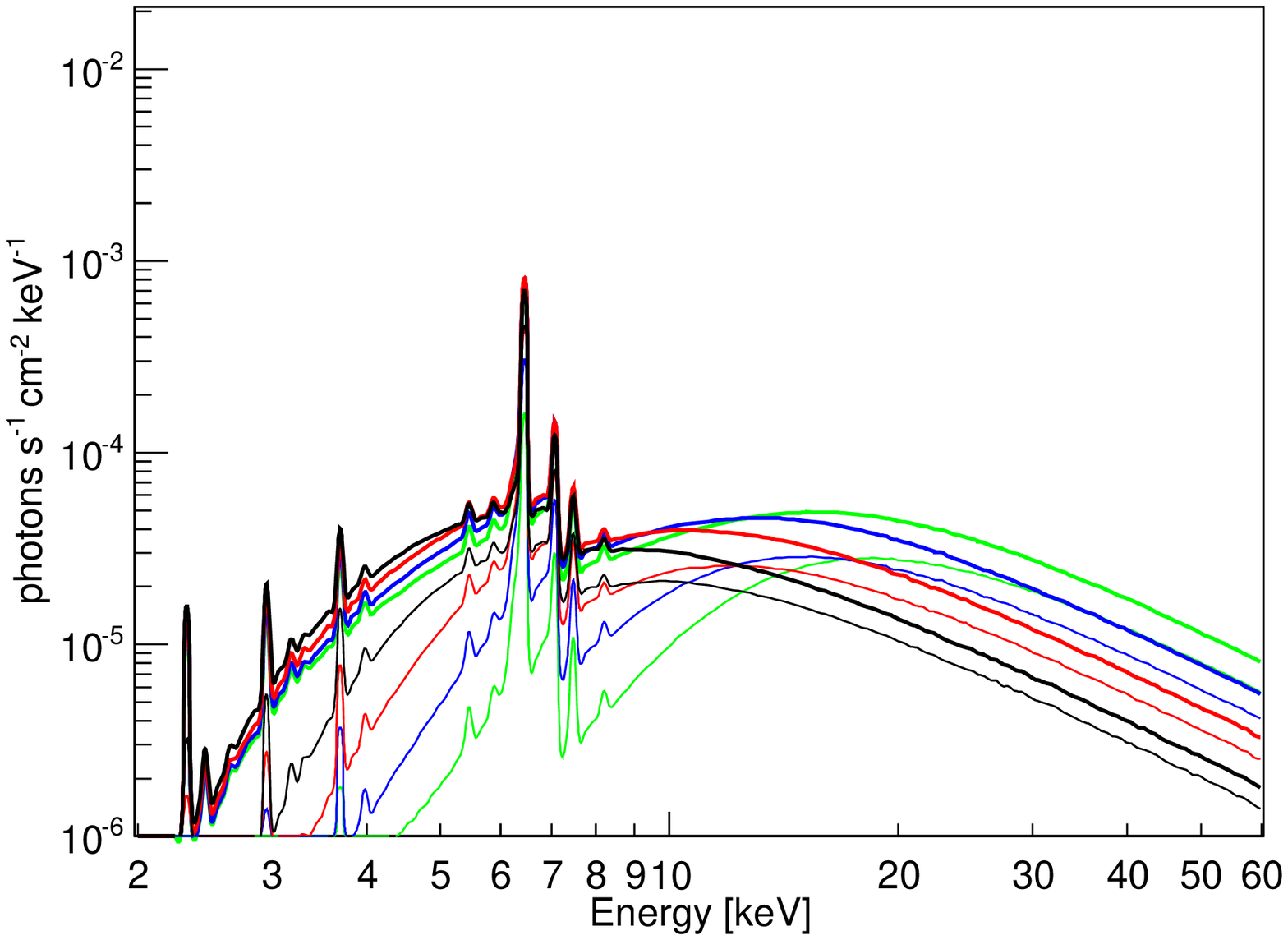}
\includegraphics[width=8.5cm]{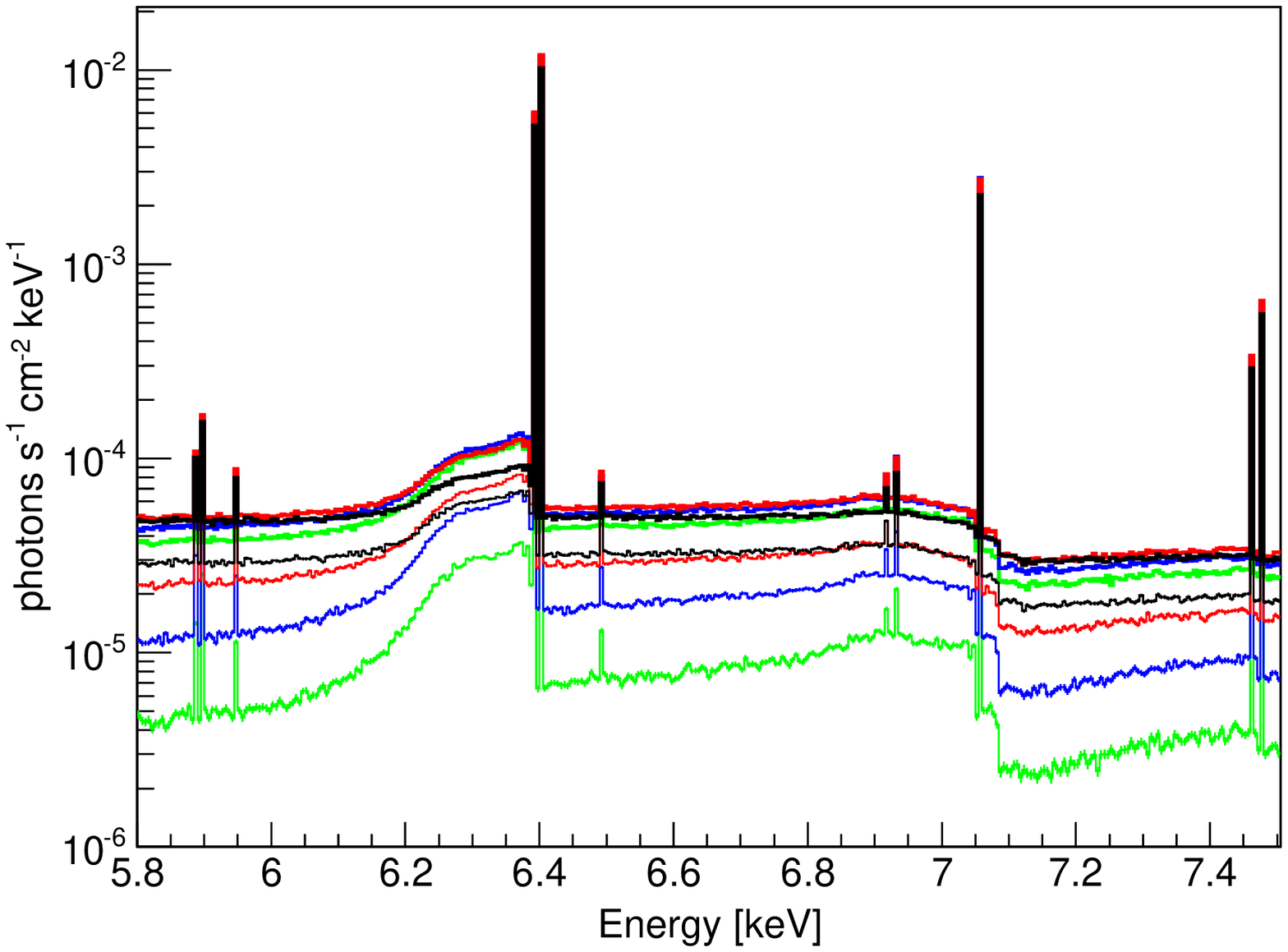}
\caption{Spectra of the reflected emissions for the cloud masses of $2.5\times 10^5M_\odot$ (Model 1; black), $5.0\times 10^5M_\odot$ (Model 2; red), $1.0\times 10^6M_\odot$ (Model 3; blue), and $2.0\times 10^6M_\odot$ (Model 4; green). The thick solid lines are spectra at $t=303\ \mathrm{yr}$ (the brightest moment) and the thin solid lines are spectra at $t=318\ \mathrm{yr}$ (fading phase). The cloud position is at $y=0\ \mathrm{pc}$. Left: wide band spectra. Right: enlarged around the iron complex with a bin width of 5~eV. Several fluorescent lines are seen: Mn K$\alpha_1$(5.90 keV), Cr K$\beta$ (5.95 keV), Fe K$\alpha_1$ (6.40 keV), Mn K$\beta$ (6.49 keV), Co K$\alpha_1$ (6.93 keV), Fe K$\beta$ (7.06 keV), and Ni K$\alpha_1$ (7.48 keV).}
\label{fig:spec_90deg}
\end{center}
\end{figure}

In addition to an image of the reflected emission, its spectrum contains important information about the cloud and the flare.
Figure~\ref{fig:spec_90deg} shows spectra extracted from the simulation results for the models with different masses (Models 1--4).
While the assumed incident spectrum is a power law with a photon index of 1.8 absorbed by the outer diffuse region ($N_\mathrm{H}=6\times 10^{22}\ \mathrm{cm}^{-2}$), absorption due to the interstellar medium is not considered here.
Since scattered hard X-rays above 20 keV do not suffer from absorption significantly, their flux increases with the cloud mass.
In contrast to the straightforward behavior of hard X-rays, dependence of the iron line on the cloud mass is complicated due to heavy absorption in dense clouds.
In addition to the iron lines, fluorescence lines of other heavy elements (e.g. Cr, Mn, Co, Ni) can be observed with high energy resolution.
Spectral evolutions are considerably different in each cloud mass since the absorption plays an important role in the fading phase of the reflected emission.
A more massive cloud displays faster decay, enhancing the line equivalent width and the Compton shoulder.

In order to evaluate the spectra quantitatively, we extracted the hard-X-ray flux (20--60 keV), the iron line flux, the equivalent width of the iron line, and the shoulder-to-peak ratio of the Compton shoulder as a function of time.
The iron line was divided into a peak and a shoulder.
The peak flux and shoulder flux are integrated over energy ranges between 6.0 keV and 6.38 keV, and in a range between 6.38 keV and 6.42 keV, respectively.
A continuum is evaluated by fitting to a power law over a range between 5.5 keV and 6.8 keV without several fluorescent line energies, and is then subtracted from both fluxes.
The equivalent width and the shoulder-to-peak ratio were calculated by using these values obtained.

\begin{figure}[htbp]
\begin{center}
\includegraphics[width=9cm]{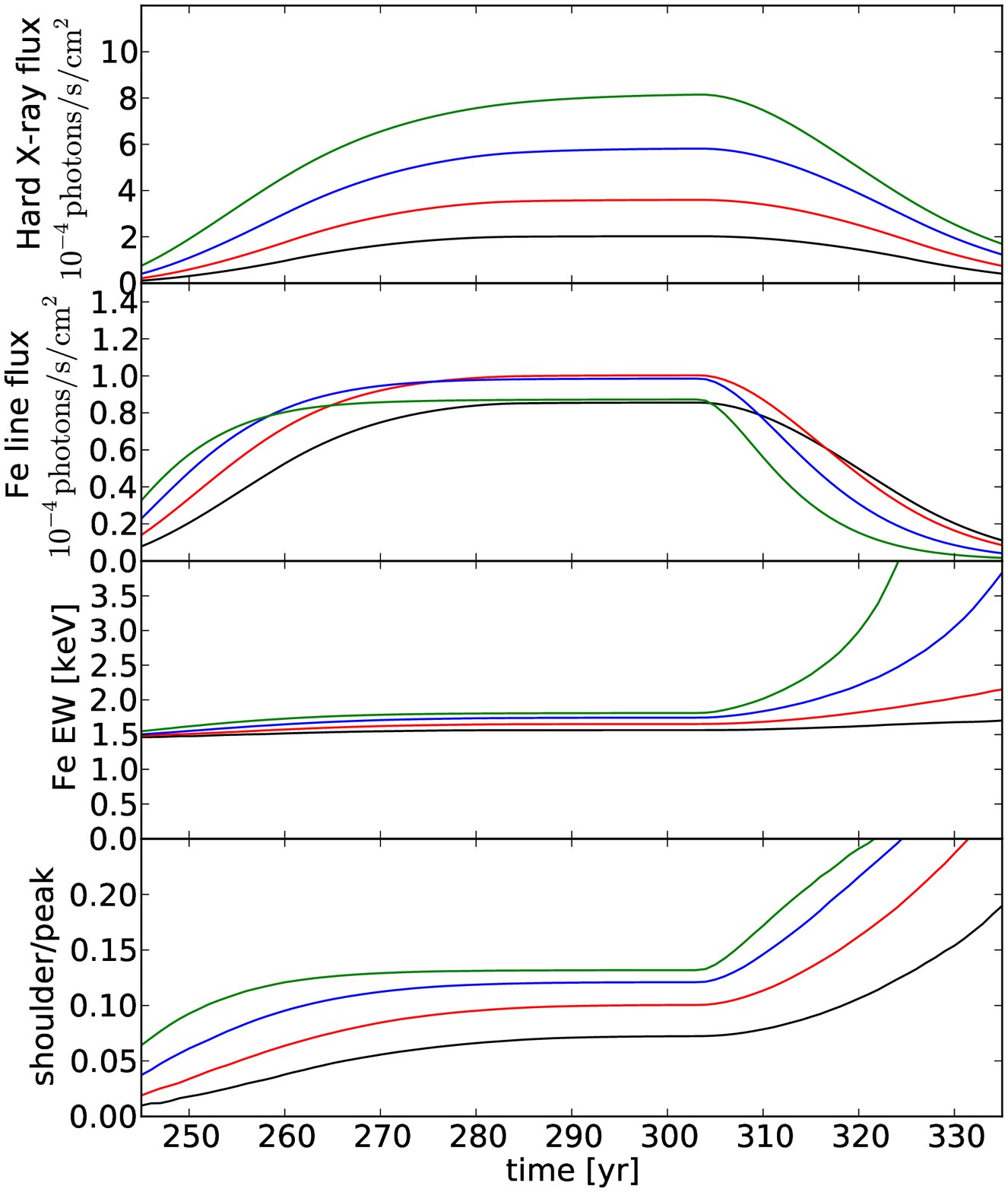}
\caption{Hard X-ray flux in a range of 20--60 keV (top), iron line flux (second), equivalent width in unit of keV (third), and shoulder-to-peak ratio (bottom) of the Fe K$\alpha$ line as a function of time for cloud masses of $2.5\times 10^5M_\odot$ (Model 1; black), $5.0\times 10^5M_\odot$ (Model 2; red), $1.0\times 10^6M_\odot$ (Model 3; blue), and $2.0\times 10^6M_\odot$ (Model 4; green). The flux unit is $10^{-4}$ photons $\mathrm{s}^{-1}\mathrm{cm}^{-2}$. The cloud position is at $y=0$ pc.}
\label{fig:lc_90deg}
\end{center}
\end{figure}

\begin{figure}[htbp]
\begin{center}
\includegraphics[width=9cm]{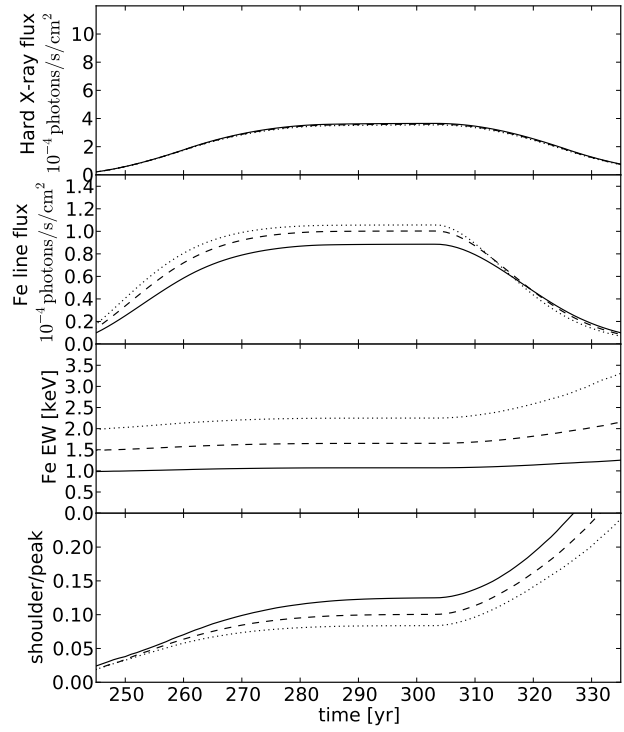}
\caption{Same as Fig.~\ref{fig:lc_90deg} but for different metal abundances of 1.0 (Model 7; solid), 1.5 (Model 2; dashed), and 2.0 (Model 8; dotted).}
\label{fig:lc_abund_90deg}
\end{center}
\end{figure}

Figure~\ref{fig:lc_90deg} shows the spectral parameters as a function of time elapsed since the end of the Sgr A* flare.
After $t=303\ \rm{yr}$, the hard-X-ray and iron line fluxes gradually decrease while the equivalent width and shoulder-to-peak ratio increase with time.
Although the iron line flux is almost independent of the cloud mass in the brightest phase, its decay speed depends on the mass because of absorption.
The equivalent width of the iron line keeps constant and does not depend on the cloud mass when the whole cloud is illuminated.
In the fading phase, however, its variation significantly depends on the mass; it shows a rapid increase in the dense clouds.
One reason for this is a difference in photoelectric absorption probability \citep{Sunyaev:1998}.
In a dense cloud, continuum photons around 6 keV are largely and rapidly absorbed though hard X-rays above 10 keV, which are responsible for producing the iron line, are not so much.
As a result, a large equivalent width of the iron line appears.
Since the Compton shoulder is generated via scattering of the iron fluorescence, it is also a good probe of the Thomson depth of the cloud or the total mass of the cloud.
The equivalent width and the shoulder-to-peak ratio also depend on the metal abundance in the cloud, as shown in Fig.~\ref{fig:lc_abund_90deg}.
The equivalent width is nearly proportional to the metal abundance.
A large metal abundance increases photoabsorption probability, suppressing scatterings to generate the shoulder. 
Note that the metal abundance controls amounts of all metals, not only iron; changing only the iron abundance could bring different results.

\begin{figure}[htbp]
\begin{center}
\includegraphics[width=9cm]{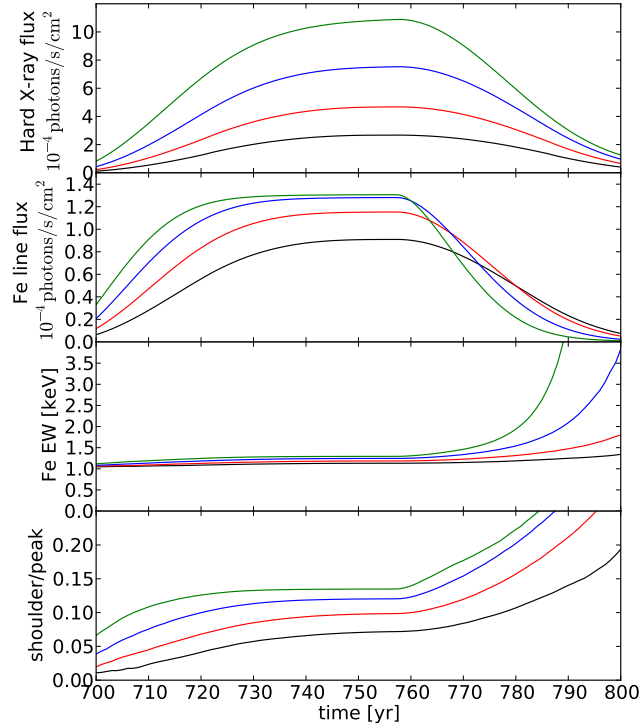}
\caption{Same as Fig.~\ref{fig:lc_90deg} but for the cloud position of $y=+100$ pc.}
\label{fig:lc_45deg}
\end{center}
\end{figure}

\begin{figure}[htbp]
\begin{center}
\includegraphics[width=9cm]{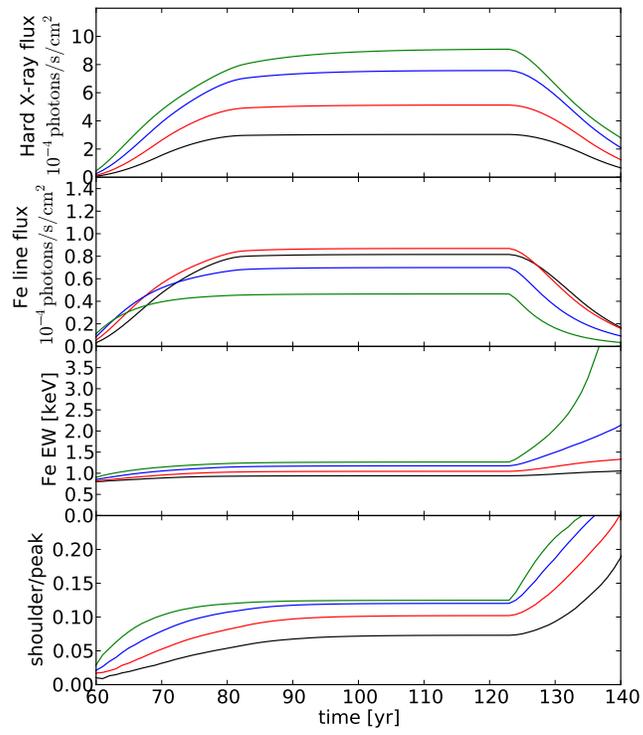}
\caption{Same as Fig.~\ref{fig:lc_90deg} but for the cloud position of $y=-100$ pc.}
\label{fig:lc_135deg}
\end{center}
\end{figure}

For different line-of-sight positions of the cloud (Fig.~\ref{fig:lc_45deg}, \ref{fig:lc_135deg}), the dependences of the iron line flux on the cloud mass are different from the cloud at $y=0$ pc, while other parameters behave in the same way.
Since the flux dependence is a monotonic function of the cloud mass, these behaviors of the iron line are somewhat easier to interpret than those of the cloud at $y=0$ pc.
For the front-illuminated cloud at $y=+100$ pc, it is an increasing function because it depends simply on the total mass of matter that generates fluorescence.
However, the line flux is saturated at $M_\text{cloud}\sim 1\times 10^6 M_\odot$.
For the back-illuminated cloud at $y=-100$ pc, on the other hand, the dependence is a decreasing function due to heavy absorption in a dense cloud.
Moreover, the line-of-sight position makes a difference to the decay time of the hard-X-ray emission without any significant biases of the cloud structure.
In contrast, the time evolution of the iron line flux depends not only on the position but also strongly on the amount of absorbing matter.
The decay time of the hard X-rays to decrease by half from the brightest moment is 13, 23, or 28 years for the different positions along the line of sight of $y=-100$, $0$, or $+100$ pc, respectively.

\section{Discussion}\label{sec:discussion}

In the previous section, we describe images and spectra of X-ray reflection from Sgr B2 obtained by the simulation.
By comparing observations with the simulation results, we can in principle constrain physical properties of both the illuminated clouds and the illuminating flare of Sgr A*.
Since there is no available data of high-resolution spectra and hard-X-ray morphologies, it is beyond the scope of this paper to obtain observational constraints.
Future data with a high signal-to-noise ratio would require more simulation runs with tuned model parameters.
Furthermore, it would be necessary to assume more realistic geometries that would not have spherical symmetry.
In these cases, our simulation framework allows us to build a more likely geometry of the cloud and to calculate emissions from it.

Although the XRN model predicts that X-ray emissions from Sgr B2 will get very dim in 10--20 years, it is fortunate that we are able to observe the response of the cloud to the incident X-rays from Sgr A* as a time variability of its morphology and spectrum.
The response provides us with information about the structure and position of the cloud, as described in the previous section.
The concave shape of the iron line emission from Sgr B2 obtained with {\it Chandra} \citep{Murakami:2001} suggests that the cloud position along the line of sight is not behind the GC.
Moreover, the fast decay time of $\sim$8 yr obtained with the {\it INTEGRAL} hard X-ray observations \citep{Terrier:2010}, which seems comparable with  that of the iron line \citep{Inui:2009}, also supports a near-side position of the cloud.
This is consistent with the radio trigonometric parallax and the X-ray absorption measurements (mentioned in \S\ref{subsec:geometry}).
A more precise estimate will be obtained by future measurements of the apparent velocity of the rear end of the light front, which is determined by the relative position of the cloud to the observer and the illuminating source.
The degree of absorption estimated by the slight distortion of the observed morphology of the iron line suggests that the mass of the cloud's envelope is smaller than the value of $\sim1\times 10^6 M_\odot$ estimated by radio observations \citep{Hasegawa:1994}.
This estimation could be inaccurate if the cloud had a more complex geometry such as a non-spherical or highly clumpy structures.

It is important to check the feasibility of the observations discussed in this paper with future instruments.
Here we describe brief prospects of the hard-X-ray imaging and high-resolution spectroscopy for {\it ASTRO-H}.
The diameter of the Sgr B2 envelope which we investigated with simulations, 10~pc, corresponds to $4'.3$ in an apparent angular size.
An angular resolution of $1.7'$ (half-power diameter or HPD) of the hard X-ray imaging system (HXI) onboard {\it ASTRO-H} allows us to pick up a few points in a spatial profile of hard-X-ray brightness across the cloud.
Despite the inadequate resolution to trace the fine structure of the cloud, the resolution is sufficient to constrain the total cloud mass and position.
Moreover, {\it NuStar} will have an angular resolution of $43''$ (HPD) or $7''.5$ (FWHM) to resolve the cloud structure in the hard-X-ray band.

\begin{figure}[htbp]
\begin{center}
\includegraphics[width=8cm]{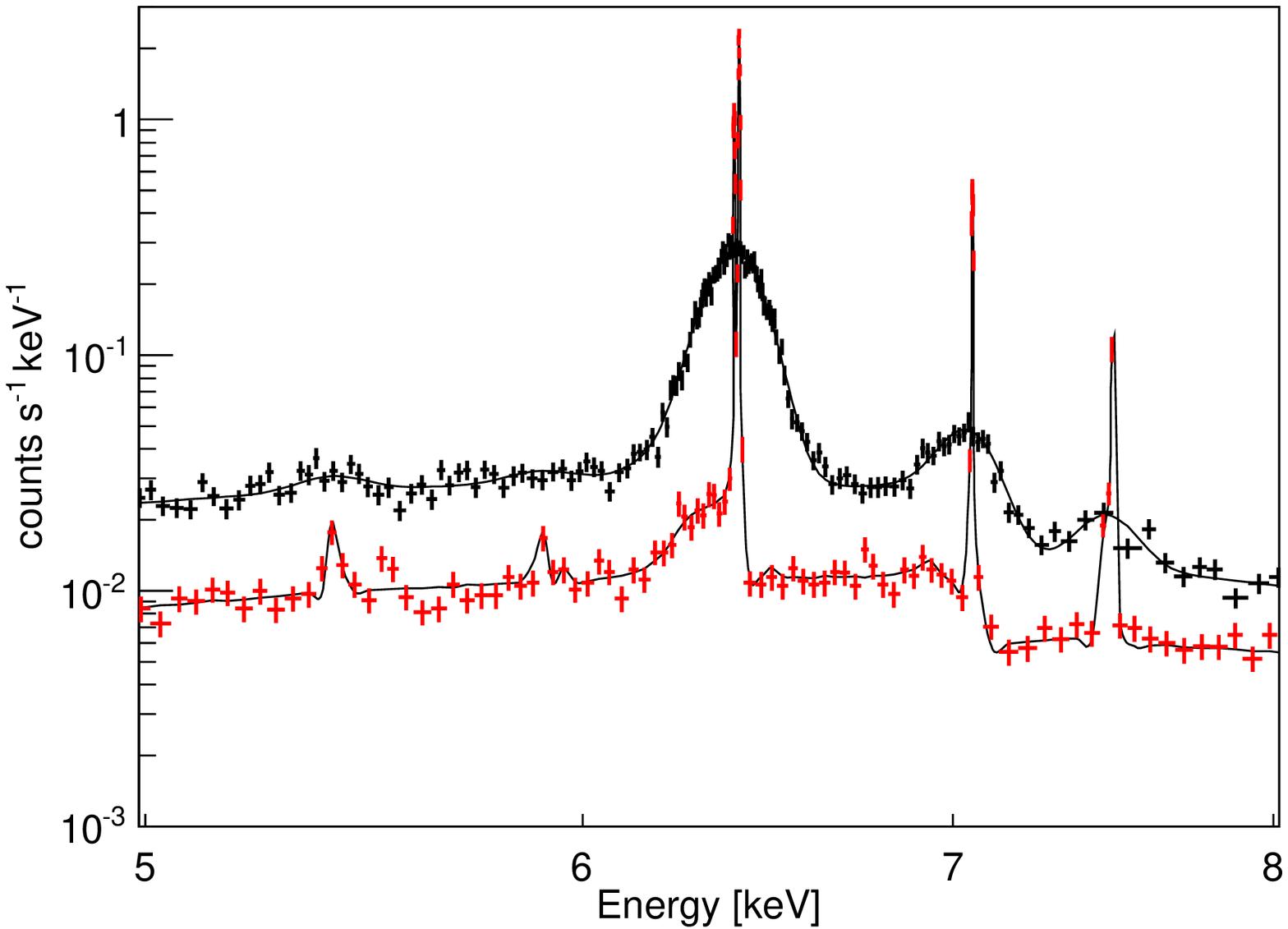}
\includegraphics[width=8cm]{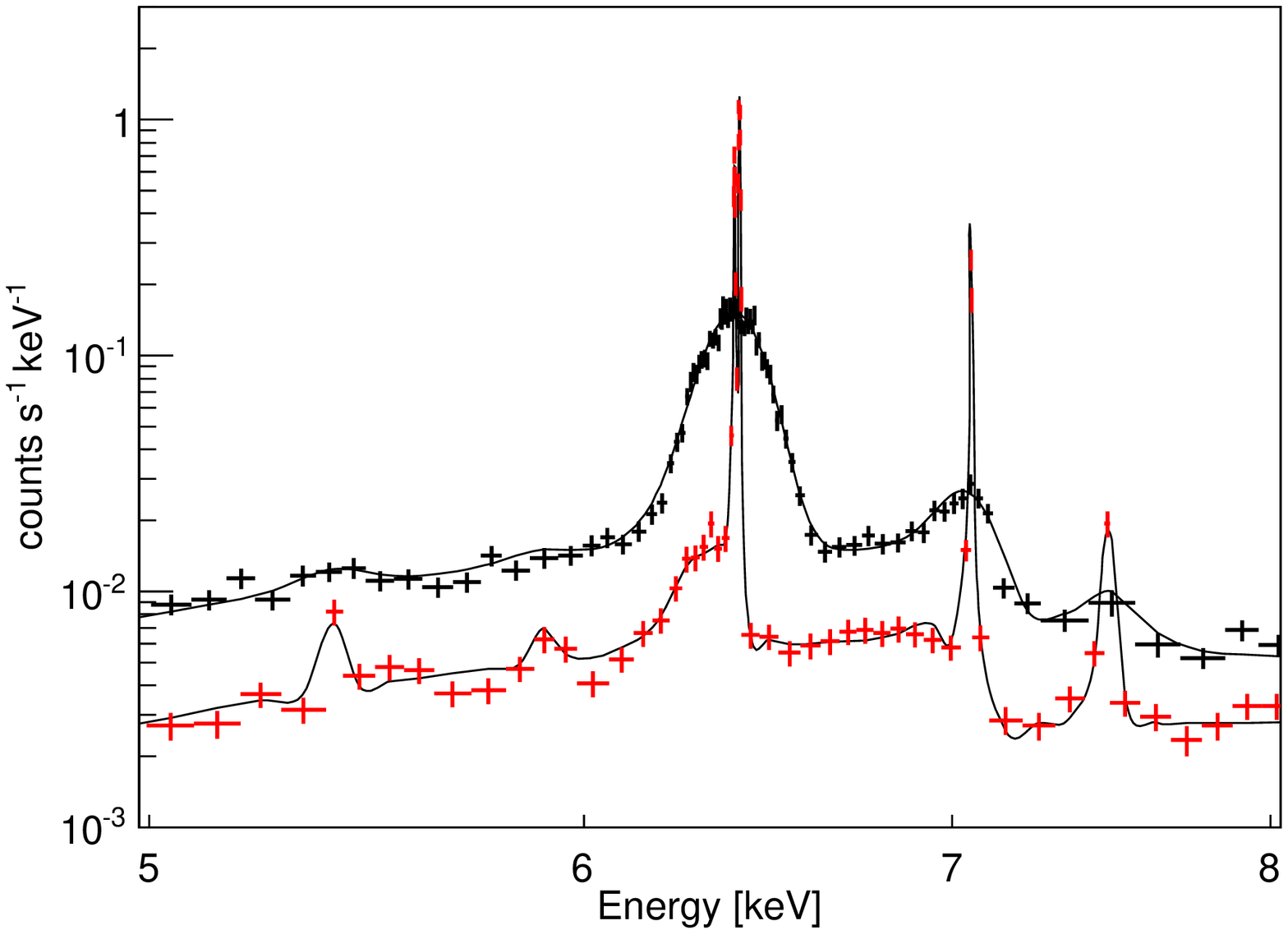}
\caption{Observation simulations of Sgr B2 based on Model 2 ($y=0$ pc) with {\it ASTRO-H} SXS (red points) and {\it Suzaku} XIS 0+3 (black points) for an exposure time of 200 ks. The left panel shows spectra at $t=303$ yr (the brightest moment) and the right panel at $t=318$ yr (fading phase). The simulated data are represented in points with error bars showing 1$\sigma$ statistical errors, superposed on solid lines representing the model convolved with the detector response. The data are re-binned for clarity. Besides iron K$\alpha$ at 6.4 keV, iron K$\beta$ 7.1 keV and nickel K$\alpha$ at 7.5 keV are clearly detected.}
\label{fig:observation}
\end{center}
\end{figure}

For high-resolution spectroscopy, we present observation simulations of Sgr B2 with the SXS onboard {\it ASTRO-H} for an exposure time of 200 ksec in Fig.~\ref{fig:observation}.
To demonstrate the advantage of the SXS over conventional observations, the same simulations with the {\it Suzaku} observatory, which has the current best spectral performance for XRNe studies, are superposed.
These predictions are based on the cloud of Model 2 positioned at $y=0$ pc at the brightest moment ($t=303$ yr) and 15 years later ($t=318$ yr).
Since the angular size of the Sgr B2 cloud may be larger than the field of view of the SXS ($3'\times3'$), we assumed that the cloud has a brightness averaged over the cloud size of 10~pc and has uniform brightness distribution in the entire field of view for the SXS simulations.
The energy resolution of the SXS is sufficient to resolve the fine structure of the iron K$\alpha$ line including its Compton shoulder at 6.4 keV, iron K$\beta$ at 7.1 keV, and fluorescent lines of less abundant metals, overcoming the soft detector response of current X-ray instruments.HM
The exposure time of 200 ksec is adequate to evaluate the line properties and continuum levels, though it depends on the actual flux of the Sgr B2 emission which will decrease.
In real observations of Sgr B2, the spectrum should be contaminated by hot plasma emissions of a temperature of 6.5 keV \citep{Koyama:2008, Ryu:2009} composed of a continuum, He$\alpha$ lines of FeXXV at 6.7 keV, and Ly$\alpha$ lines of FeXXVI at 6.9 keV.
For accurate estimation of the contamination level, the excellent energy resolution of the SXS will be essential to measuring the plasma temperature, which determines the continuum emission via bremsstrahlung.
This estimate of the hot plasma contamination will be applicable to X-ray observations with current instruments to obtain more precise values of the equivalent width of the iron line when we safely assume the invariable flux of the thermal emission.

In addition to the Sgr B2 complex, other molecular clouds in the GC regions are also valuable XRNe for studying the Sgr A* activity and the structure of the molecular clouds.
The fact that a part of the molecular clouds around Sgr A display time variability of the iron fluorescence provides strong evidence of the past activity of Sgr A* and information on its history \citep{Ponti:2010}.
Our model of XRNe enables us to constrain the positions and physical properties such as total masses of clouds in the GC regions.
Based on the constraints thus obtained, we could construct a consistent view of the structure of the GC molecular zone and the history of the activity of the Sgr A* supermassive black hole from a few thousand years ago to the recent past several decades.
Moreover, there are giant molecular clouds that are faint in X-ray near Sgr A*.
Future missions with a large effective area will be able to investigate such clouds by means of a Compton shoulder which is a remnant of past illumination.

\section{Conclusions}\label{sec:conclusion}

We have developed a new calculation framework of X-ray reflection from giant molecular clouds in the GC region (e.g.~Sgr B2) based on Monte Carlo simulations.
Our model carefully treats radiative transfer including multiple interactions in the complicated geometry of a cloud.
By utilizing this simulation framework, we investigated X-ray diagnostics of XRNe aimed at hard-X-ray imaging and high-resolution spectroscopy for which future missions such as {\it ASTRO-H} will open a window.
Scattered hard X-rays from the cloud show significantly different morphologies from the iron fluorescence due to their large penetrating power into a dense cloud.
Thus, hard-X-ray imaging will provide new important information on the structure of the illuminating molecular clouds.
High-resolution spectroscopy also brings a quantitative approach to constraining the physical properties of the cloud such as the total mass and the chemical compositions of metals, as well as the luminosity of the illuminating source.
A large-scale view of X-ray reflection from molecular clouds in the GC region will reveal the structure of the GC molecular zone and the history of the activity of the central black hole in our Galaxy.

\acknowledgments

We thank Eugine Churazov for providing us with data concerning differential cross sections of scattering by He atom.
H.~Odaka is supported by research fellowships of the Japan Society for the Promotion of Science for Young Scientists.
This work was supported in part by Global COE Program (Global Center of Excellence for Physical Sciences Frontier), MEXT, Japan.

\appendix

\section{Framework of Monte Carlo simulation}\label{app:framework}
 
\subsection{Weighting Method}\label{subsec:convolution}

If we assumed that radiation did not affect the surrounding medium, X-ray reprocessing (fluorescence or scattering) in the system would be regarded as a linear response of the system.
The linear nature of the system greatly simplifies the problem because the final results are obtained by convolving the response function, which describes how the system changes the initial photon into the final state of the photon.
In other words, the problem is reduced to obtaining a Green's function of a linear operator which describes the system.
X-ray reflection by cold matter can be treated in this manner.

Thus, the purpose of the Monte Carlo simulation is to calculate a conditional probability density of emitting a photon with an energy $E_1$ and direction $\bm{\Omega}_1$ at time $t_1$ and position $\bm{x}_1$ under initial conditions of a photon emitting with an energy $E_0$ and direction $\bm{\Omega}_0$ at time $t_0$ and position $\bm{x}_0$.
Figure~\ref{fig:concept} shows a schematic concept of the simulation.
In each Monte Carlo trial, we track a photon from certain initial conditions until the photon escapes from the system, recording the initial condition and the conditions of photons at the last interaction before escaping.
After $M$ trials of Monte Carlo simulation, the conditional probability density can be written as
\begin{equation}
p(E_1, \bm{\Omega}_1, t_1, \bm{x}_1|E_0, \bm{\Omega}_0, t_0, \bm{x}_0) = \frac{g(E_1, \bm{\Omega}_1, t_1, \bm{x}_1|E_0, \bm{\Omega}_0, t_0, \bm{x}_0)}{\beta (E_0, \bm{\Omega}_0, t_0, \bm{x}_0)},
\end{equation}
where $g$ denotes the number of events that have an initial condition of $X_0=(E_0, \bm{\Omega}_0, t_0, \bm{x}_0)$ and a final condition of $X_1=(E_1, \bm{\Omega}_1, t_1, \bm{x}_1)$, and $\beta$ denotes a distribution function of initial conditions of the simulation, which is normalized as
\begin{equation}
\int \beta(X_0) dX_0 = M.
\end{equation}
Then, we can write a distribution function of emitting photons in the system as
\begin{equation}\label{eq:distribution_func_emission}
\begin{split}
f(X_1)&=\int p(X_1|X_0) \alpha(X_0) dX_0=\int g(X_1|X_0)\frac{\alpha(X_0)}{\beta(X_0)}dX_0,
\end{split}
\end{equation}
where $\alpha(X_0)$ is a distribution function of the initial photons that are emitted by the source, or simply the source function.
The simulation output is a list of the escaping photons with information on the initial and final conditions.
For the list, $f(X_1)$ is calculated by
\begin{gather}
f(X_1)=\sum_{i=1}^{M} w_i(X_1', X_0),\\
\label{eq:event_weight}
w_i(X_1', X_0) = \left\{ \begin{array}{ll}
\dfrac{\alpha(X_0)}{\beta(X_0)} & (X_1' = X_1) \\
0 & (X_1'\neq X_1)
\end{array} \right.
\end{gather}
Here, $w_i$ is a weight of each event and the summation is performed over all Monte Carlo events.
When the geometry of the system is static or has a symmetry of time translation, it is possible to fix the initial time $t_0$ in the simulation to zero;
then Eq.~(\ref{eq:distribution_func_emission}) is reduced to
\begin{equation}
f(E_1, \bm{\Omega}_1, t_1, \bm{x}_1)=\int g(E_1, \bm{\Omega}_1, t_1-t_0, \bm{x}_1|E_0, \bm{\Omega}_0, 0, \bm{x}_0)\frac{\alpha(E_0, \bm{\Omega}_0, t_0, \bm{x}_0)}{\beta(E_0, \bm{\Omega}_0, 0, \bm{x}_0)}dE_0 d\bm{\Omega}_0 dt_0 d\bm{x}_0.
\end{equation}

To simulate observations, we must consider time differences of light propagation to an observer from different positions.
A distant observer directed toward an unit vector $\bm{r}$ observes the emission at $t_\text{obs}= t_1-\bm{x}_1\cdot \bm{r}/c$, where $c$ is speed of light and the origin of $t_\text{obs}$ is determined so that an emission at $t_1=0$ and $\bm{x}_1=\bm{0}$ is observed at $t_\text{obs}=0$ through direct propagation (without any interactions).
Events are selected to obtain observational spectra or images if the direction agrees with $\bm{r}$ within a certain small tolerance.
The weight of the selected event is calculated by Eq.~(\ref{eq:event_weight}) in which $t_1=t_\text{obs}+\bm{x}_1\cdot \bm{r}/c$ is substituted.
If the system has spatial symmetry such as spherical or axial symmetry, we can extend the range of the direction selection to reduce statistical uncertainties unless the symmetry breaks.

\subsection{Framework Design}

\begin{figure}[htbp]
\begin{center}
\includegraphics[width=12cm]{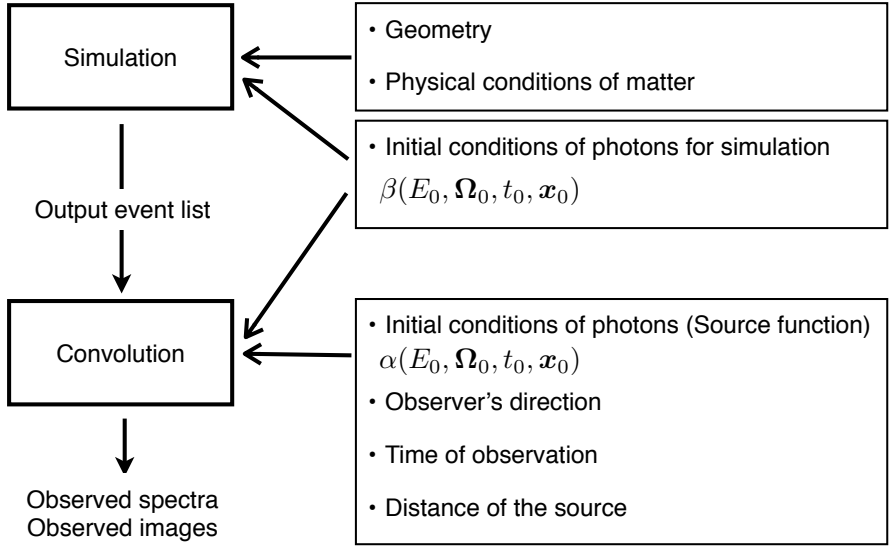}
\caption{Outline of our simulation framework}
\label{fig:framework1}
\end{center}
\end{figure}

Our simulation framework, MONACO, provides methods for the Monte Carlo simulation and for the convolution to generate observed spectra or images (\S\ref{subsec:convolution}).
An outline of the calculation to obtain observational data is shown in Fig.~\ref{fig:framework1}.
The first part uses Monte Carlo simulations to calculate the probability of obtaining certain final conditions, given certain initial conditions.
This part requires information about the initial conditions of the photon and the geometry including physical conditions of matter as input.
The output here is a list entry for an event that has information about the initial state and the final state (emission).
If a photon does not experience an interaction with matter, the final state is identical to the initial state.

Using the event list generated by the simulation part as an input, the second part, referred to ``convolution'' in Fig.~\ref{fig:framework1}, generates simulated observational data such as spectra or images by performing the convolution with the source function described in \S\ref{subsec:convolution}.
The output here is the final results, which can be compared with real observations.
The list input can be a combination of different event lists to reduce statistical uncertainties or to extend the ranges of initial conditions.
To perform the convolution by Eq.~(\ref{eq:event_weight}), this part requires as input the initial conditions of photons both for the simulation and for the real source (source function).
It also requires information on the observer; direction, time, distance and miscellaneous parameters.

The first part (simulation part) performs the photon-tracking simulation including treatment of X-ray reprocessing.
This part is an extended and improved version of original Monte Carlo codes developed by \citet{Watanabe:2003, Watanabe:2006}.
We use the Geant4 simulation toolkit library \citep{Agostinelli:2003, Allison:2006} only for treatment of complicated geometry while physical processes are our original implementations.
The simulation framework consists of a manager and four user-defined modules.
The manager, which is provided by the Geant4 library, handles communications between the modules and treats particle tracking based on Monte Carlo simulations and geometry navigation.
The user-defined modules are for geometry, physical processes, primary generation, and data acquisition.
The geometry module describes the geometry of the astrophysical object to be simulated as well as physical conditions of the matter.
The module for physical processes organizes X-ray interactions with matter.
The primary generator describes how to create initial photons to be tracked.
For example, a point-like source is specified by a position vector of the source and an energy spectrum such as a power law with a photon index and an energy range; in each Monte Carlo trial, the direction and energy of the primary photon are sampled from an isotropic distribution and the power-law distribution, respectively.
The data acquisition module forms an event list of the simulation data which contains information on the initial and final states of the photon and saves it to a file.

\section{Scattering by Bound Electrons in Neutral Matter}\label{app:process_scattering}

\citet{Sunyaev:1996} give a detailed review of scattering by bound electrons in neutral matter such as molecular clouds.
Here we summarize physical implementation of this scattering process.
This process can be divided into three channels according to the final state of the target electron:
\begin{equation}\label{eq:scattering_bound}
\gamma+\mathrm{X}_i \to \left\{ \begin{array}{ll}
\gamma_1+\mathrm{X}_{i} & \text{(ground state---Rayleigh scattering)} \\
\gamma_1+\mathrm{X}^*_{i} & \text{(excited state---Raman scattering)} \\
\gamma_1+\mathrm{X}_{i-1} + \mathrm{e}^{-} & \text{(free state---Compton scattering)}
\end{array}\right.
\end{equation}
Assume that a photon is scattered by an electron bound by an atom.
Through this scattering, the quantum state of the electron changes from an initial state $|i\rangle$ to a final state $|f\rangle$.
The doubly differential cross sections for atomic hydrogen are given by,
\begin{gather}
\frac{d^2\sigma}{d\Omega dh\nu_2}=
\left(\frac{e^2}{m_\mathrm{e}c^2}\right)^2 \left(\frac{\nu_2}{\nu_1}\right) (\bm{e}_1\bm{e}_2)^2 
\sum_f \left| \langle f|e^{i\bm{\chi}\bm{r}}|i\rangle \right|^2
\delta(E_f-E_i-\Delta h\nu), \\
h\nu_1 + E_i = h\nu_2 + E_f, \quad \Delta h\nu = h\nu_1 - h\nu_2, \\
\bm{q}=\bm{k}_1 - \bm{k}_2, \quad \bm{\chi}=\bm{q}/\hbar.
\end{gather}
Here, $E_i$ and $E_f$ are the initial and the final energies of the electron;
$h\nu_1$ and $\bm{e}_1$ are the energy and unit vector of polarization of the initial photon, respectively;
$h\nu_2$ and $\bm{e}_2$ are those of the final photon.
The momentum transfer through this scattering is denoted by $\bm{q}$; $\bm{k}_1$ and $\bm{k}_2$ are the initial and the final momenta of the photon.

For atomic hydrogen, we have analytic solutions of the electron wave function. Thus, the differential cross sections of the three channels can be calculated by analytic formulae.
For Rayleigh scattering, the differential cross section is 
\begin{equation}\label{eq:diff_cs_rayleigh}
\frac{d\sigma}{d\Omega}=
\left( \frac{d\sigma}{d\Omega} \right)_\text{Th}
\left[ 1+ \left( \frac{1}{2}qa \right)^2 \right]^{-4},
\end{equation}
where
\begin{equation}
\left( \frac{d\sigma}{d\Omega} \right)_\text{Th}=
\frac{1}{2}r_\mathrm{e}{}^2 (1+\cos^2\theta)\left( \frac{\nu_2}{\nu_1} \right)^2,
\end{equation}
$a$ is $r_\mathrm{B}/\hbar$, and $r_\mathrm{B}=\hbar/m_\mathrm{e}c\alpha$ is the Bohr radius.
For Raman scattering to the final electron state of $n$ (principal quantum number), the differential cross section is given by
\begin{equation}
\frac{d\sigma}{d\Omega}=
\left( \frac{d\sigma}{d\Omega} \right)_\text{Th}
\frac{2^8}{3}\frac{(qa)^2}{n^3}
\left[ 3(qa)^2+ \frac{n^2-1}{n^2}\right]
\frac{\left[ (n-1)^2 / n^2 + (qa)^2 \right]^{n-3} }{\left[ (n+1)^2/n^2 + (qa)^2 \right]^{n+3} }.
\end{equation}
Substituting $n=1$ to this expression of course yields Eq.~(\ref{eq:diff_cs_rayleigh}).

Compton scattering by a bound electron makes a free electron, and the energy of the scattered photon is not determined uniquely by the scattering angle.
The doubly differential cross section is given by
\begin{equation}
\frac{d^2\sigma}{d\Omega dh\nu_2} = 
\left( \frac{d\sigma}{d\Omega} \right)_\text{Th}
\left(\frac{\nu_1}{\nu_2}\right)
H_{fi}{}^2,
\end{equation}
where
\begin{equation}
\begin{split}
H_{fi}{}^2 =& 2^8 a^2 m(1-e^{-2\pi /pa})^{-1} \exp\left[ \frac{-2}{pa}\tan^{-1}\left(\frac{2pa}{1+q^2 a^2 -p^2 a^2}\right) \right] \\
&\times \left[ q^4 a^4 + \frac{1}{3}q^2 a^2 (1+p^2 a^2) \right] \\
& \times [(q^2 a^2 + 1 - p^2 a^2)^2 + 4p^2 a^2]^{-3}, \\
p^2/2m =& -|E_b|+\Delta h\nu.
\end{split}
\end{equation}
Here, $p$ and $E_\mathrm{b}$ are the momentum of the ejected electron and the binding energy of the electron, respectively.
The energy difference of the photon should be larger than the binding energy $E_\mathrm{b}=13.6$ eV for Compton scattering.

\begin{figure}[htbp]
\begin{center}
\includegraphics[width=9cm]{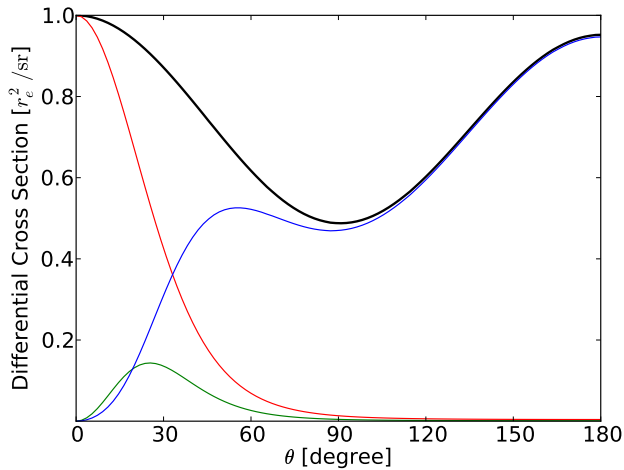}
\caption{Differential cross sections for the three scattering channels of a 6.4 keV photon. The thin lines show contributions of Rayleigh scattering (red), Raman scattering (green) and Compton scattering (blue). The thick black line is the sum of the total contributions. The values of the differential cross sections are normalized by the classical electron radius squared $r_\mathrm{e}{}^2$.}
\label{fig:dcs_hydrogen_atom}
\end{center}
\end{figure}

Figure~\ref{fig:dcs_hydrogen_atom} shows the differential cross sections for the three channels of a 6.4 keV photon, corresponding to the iron K$\alpha$ line.
For small scattering angles, Rayleigh scattering is important, and Raman and Compton scattering are suppressed since the electron could not gain enough energy to be ionized or excited from the incident photon.
On the other hand, Compton scattering becomes dominant at the large scattering angles and agrees with Compton scattering by a free rest electron described by Klein--Nishina formula.

In cold, dense matter, hydrogen exists in forms of molecules $\mathrm{H}_2$, not as atoms H.
Since the electron state of the hydrogen molecule differs from that of atomic hydrogen, it is important to consider the molecular effects on the scattering processes.
This problem under astrophysical applications is discussed by \citet{Sunyaev:1999}.
The most significant difference between a single-electron atom and a molecule is the enhancement of Rayleigh scattering due to electron coherence.
Inelastic scattering (Raman and Compton) does not differ very much from that of atomic hydrogen.
Moreover, effects of rotational and vibrational states of the molecule are negligible for the X-ray energy band.
Thus, we adopt a good approximation for molecular hydrogen proposed by \citet{Sunyaev:1999} that scattering by the molecule is identical to that by atomic hydrogen except that the cross section of Rayleigh scattering is enhanced by a factor of two per electron.

In many astrophysical situations, the second most abundant element, helium, also makes an important contribution to scattering.
Like molecular hydrogen, atomic helium has two electrons, resulting in enhancement of the coherent scattering.
In addition, helium has larger binding energy $E_\mathrm{b}=24.6$ eV and larger momentum of bound electrons than those of hydrogen.
These affect the energy profile of scattered photons via Compton scattering.
Since helium does not have analytic expressions of atomic properties, we use numerical calculations of differential cross sections provided by \citet{Vainshtein:1998}.

\begin{figure}[htbp]
\begin{center}
\includegraphics[width=7.5cm]{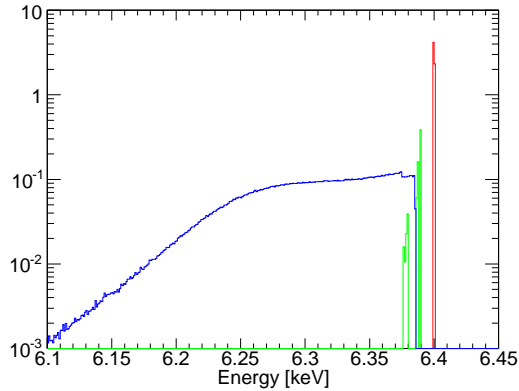}
\caption{Spectra of scattered photons of a monochromatic 6.4 keV line via Rayleigh scattering (red), Raman scattering (green), and Compton scattering (blue) in arbitrary units.}
\label{fig:spec_scattering}
\end{center}
\end{figure}

Figure~\ref{fig:spec_scattering} shows calculated spectra of scattered photons of a monochromatic 6.4 keV line via three different channels.
Rayleigh scattering does not change the photon energy.
There is a gap with a width of $\sim$10.2 eV that corresponds to the first excitation energy of hydrogen.
Compton scattering is permitted for the scattered energy lower than $E_\text{line}-E_\mathrm{b}=6.4-0.0136=6.3864$ keV due to hydrogen's binding effects.
A discontinuous edge at 6.375 keV is due to the same effect by helium.
This edge can be used for measurement of helium abundance for future large-area missions \citep{Vainshtein:1998}.
The profile of the Compton scattering is smeared by the initial momentum of a target electron bound to an atom, displaying a long low-energy tail.
If the electron were free and had zero momentum, the profile would have a sharp edge at $E=6.24 $keV, corresponding to back scattering ($\theta=180^\circ$).

\bibliography{gc_simulation_final}

\end{document}